\documentclass[preprint,12pt]{elsarticle}%
\usepackage{graphicx}
\usepackage{amssymb}
\usepackage{amsfonts}
\usepackage{amsmath}
\usepackage{color}%
\providecommand{\U}[1]{\protect\rule{.1in}{.1in}}
\providecommand{\U}[1]{\protect\rule{.1in}{.1in}}
\biboptions{numbers,sort&compress}
\journal{Journal of Sound and Vibration}

\begin{document}

\begin{abstract}
Many engineering structures are composed of weakly coupled sectors assembled in a cyclic and ideally symmetric configuration, which can be simplified as forced Duffing oscillators. In this
paper, we study the emergence of localized states in the weakly nonlinear
regime.  We show that multiple spatially localized solutions may exist, and the
resulting bifurcation diagram strongly resembles the snaking pattern observed
in a variety of fields in physics, such as optics and fluid dynamics. Moreover,
in the transition from the linear to the nonlinear behaviour isolated branches
of solutions are identified. Localization is caused by the hardening effect
introduced by the nonlinear stiffness, and occurs at large excitation levels.
Contrary to the case of mistuning, the presented localization mechanism is
triggered by the nonlinearities and arises in perfectly homogeneous systems.

\end{abstract}

\begin{frontmatter}%
\title{Multistability and localization in forced cyclic symmetric structures modelled by weakly-coupled Duffing oscillators}%
\author{A. Papangelo$^{(1,2,*)}$, F. Fontanela$^{(3)}$, A. Grolet$^{(4)}%
$, M. Ciavarella$^{(1,2)}$, N. Hoffmann$^{(2,3)}$}%
\address{$^{(1)}$Politecnico di BARI. DMMM dept. V Gentile 182, 70126 Bari, Italy}
\address{$^{(2)}%
$Hamburg University of Technology, Department of Mechanical Engineering, Am Schwarzenberg-Campus 1, 21073 Hamburg, Germany}
\address{$^{(3)}%
$Imperial College London, Exhibition Road, London SW7 2AZ , UK}
\address{$^{(4)}%
$Arts et M\'etiers ParisTech, Department of Mechanical Engineering, 8 Boulevard Louis XIV, 59000 Lille, France}
\address{$^{(*)}$email: antonio.papangelo@poliba.it }%
\end{frontmatter}

\section{Introduction}

The aerospace and aeronautical industries face an increasing effort to improve
machine performances \cite{Bartels2007}. Within this scenario, several new
designs have been proposed, some of which relying on lightweight materials
applied {to} e.g. bladed disks, satellites, and space antennas. The resulting
slender structures may undergo large deflections and nonlinear phenomena such
as energy localization may be observed.

The problem of energy localization is a well-known phenomenon within the
vibration engineering community, particularly in turbomachinery \cite{Ewins1969, newcit2}, aerospace structures \cite{newcit} but also in microelectromechanical systems \cite{Dick2008,Dick2010}. In the linear
framework, indeed, that community usually refers to the behaviour as a
mistuning problem due to its relevance for the design of bladed disks
\cite{Ewins1969,Hodges1982}. In this case, localization arises due to the
break of symmetry induced by inhomogeneities from manufacturing variability or
wear. From the modelling perspective, the capability of computing the level of
localization has been extensively addressed \cite{Whitehead1966,Castanier2006}%
, although a real solution to the problem has not been found: small changes of
the mass distribution (due to erosion or wear) during service make any clever
solution impossible, and design has to take into account worst case
scenarios involving significant amplification of vibration with respect to the
homogeneous case. Linear energy localization is known also more in general in
other research fields. For example, in the solid state physics community, it
is known as "Anderson localization", as the phenomenon was firstly studied by
Anderson \cite{Anderson1958} in the context of a diffusion wave problem, and
was a Nobel-prize discovery. Nevertheless, in complex engineering applications
the linear behaviour has to be seen as an approximation, as non-linearities
emerge in either stiffness or damping elements. In the case{ of} energy
dissipation in bladed disks, for example, vibration control usually relies on
passive components, such as frictional dampers and this induces both
non-linearities due to the contact status and to the frictional law
\cite{Sanliturk1999, Firrone2009, pap2015, wedgeshaped2016, Pesaresi1, Pesaresi2}.
Therefore, even the standard vibration control technique implemented in the
aerospace industry inherently introduces nonlinearity to the physical systems.

For aeronautical applications the so-called \textquotedblright
blisk\textquotedblright\ have been developed, where contact surfaces between
the disk and the blade root are avoided as the whole structure is built
monolithically. Due to the lack of frictional dissipation in joints, blisks suffer the lack of sufficient damping, which can expose the
structures to large deformation regimes, where geometric nonlinearities have
an important role. In this case, energy localization goes much beyond
mistuning (see e.g. \cite{Vakakis, Kerschen2009}). It is well-known, for
example, that even perfect cyclic and symmetric structures may localize energy, due to the evolution of non-similar modes or due to bifurcations, i.e. qualitative changes in system dynamics (see e.g. \cite{Grolet2012,Fontanela2017,Starosvetsky2013,Papangelo2017a,
Papangelo2017b,Papangelo2018,Bendiksen1987,Bendiksen1989,Georgiades2009,Castanier2002}%
). Nonlinear localization has been shown to appear not only in conservative
systems \cite{Papangelo2017a} but also in friction-excited chains of weakly
coupled oscillators \cite{Papangelo2017b,Papangelo2018}, where the authors
showed that the localization phenomenon is strongly related to the bistable
behaviour of the single oscillator in the chain. In the bifurcation diagram
snaking-like bifurcations appear that strongly resemble the \textquotedblright
snaking bifurcations\textquotedblright\ observed in different physics fields,
such as fluid dynamics \cite{Champneys1998, Thual1988, Beaume2011}\ and
nonlinear optics \cite{Champneys1998}.

In this paper, we study an harmonically forced cyclic and symmetric system
made of a chain of Duffing oscillators with weak nearest-neighbour elastic
coupling. The system can be seen as a minimal model for several cyclic and
symmetric structures in the weakly nonlinear regime. It will be shown that the
level of vibration localization depends on the forcing level, and it disappears in the
case of low excitation when the system tends to the linear behaviour.
Moreover, in the transition from low to high excitation, isolas of solutions
appear. When the forced response is compared with the nonlinear normal modes
of the system, it is shown that local resonances perfectly overlap with a
particular backbone curve. 

Notice that the mechanical system under study is homogeneous in space, thus the localization phenomenon occurs due to the inherent nonlinearities in the system (cubic stiffness terms) and is not related to any kind of inhomogeneity. In the last part of the paper we discuss the results obtained and point to possible further extension of this work.

\section{A lumped nonlinear model for cyclic symmetric structures}

The model under study consists of $N_{s}$ identical oscillators with mass $m$,
cyclically connected to each other by linear springs $k_{c}$, and also
attached to the ground by linear springs and viscous dampers $k_{l}$ and $c$,
respectively (Fig. 1). Each oscillator is also excited by an external force
$f_{n}(t)$ and is subjected to nonlinear forces induced by the cubic stiffness
$k_{nl}$. The physical system in Fig. 1 can be considered a minimal model for
aerospace structures, e.g. blisks, antennas, or reflectors, undergoing large deformations.

\begin{center}
\includegraphics[
height=1.7231in,
width=5.0788in
]{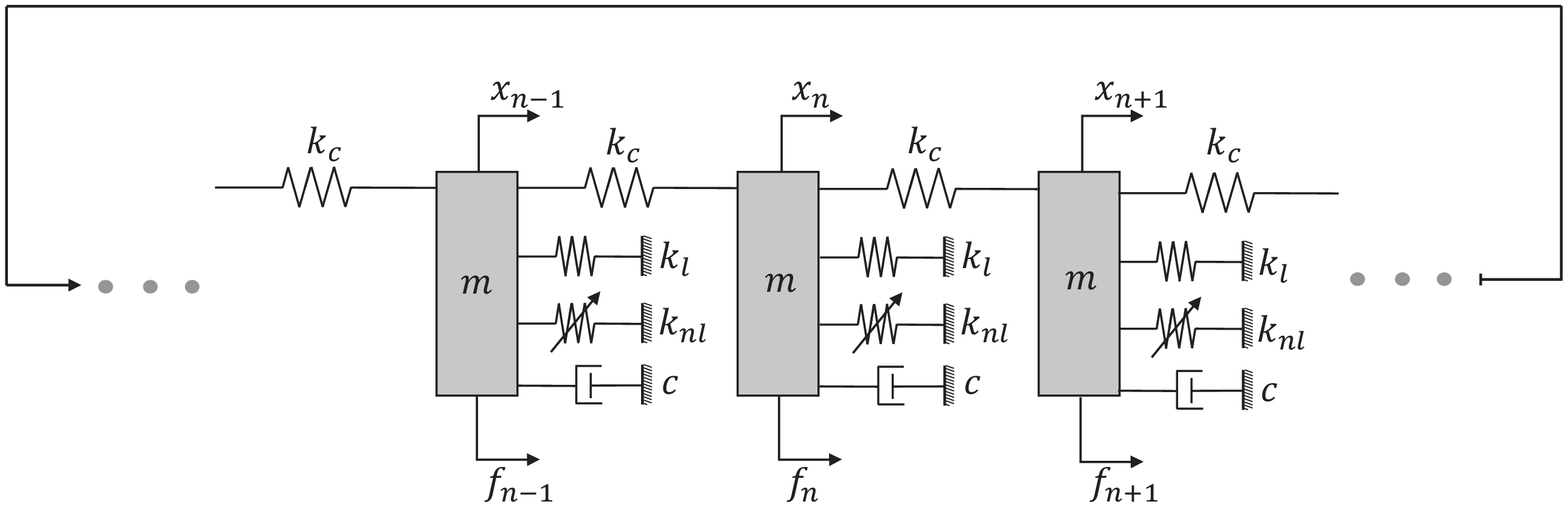}

Fig. 1 - Minimal model for a cyclic and symmetric structure undergoing large displacements
\end{center}

The equation of motion for the $n$-th oscillator is written as
\begin{equation}
m\ddot{x}_{n}+c\dot{x}_{n}+k_{l}x_{n}-k_{c}(x_{n+1}+x_{n-1}-2x_{n}%
)+k_{nl}x_{n}^{3}=f_{n}(t) \label{Eq:Mov1}%
\end{equation}
where $\ddot{x}_{n}$, $\dot{x}_{n}$, and $x_{n}$ are the acceleration,
velocity, and displacement, respectively, while $f_{n}(t)$ represents the
corresponding external force applied to the $n$-th mass. Equation
\eqref{Eq:Mov1} is rewritten, for convenience, such that
\begin{equation}
\ddot{x}_{n}+2\xi\omega_{0}\dot{x}_{n}+\omega_{0}^{2}x_{n}-\omega_{c}%
^{2}(x_{n+1}+x_{n-1}-2x_{n})+\gamma x_{n}^{3}=\frac{f_{n}(t)}{m},
\label{Eq:Mov2}%
\end{equation}
where $\omega_{0}^{2}$=$\frac{k_{l}}{m}$, $\omega_{c}^{2}$=$\frac{k_{c}}{m}$,
$\gamma$=$\frac{k_{nl}}{m}$, and $\xi$=$\frac{c}{2\sqrt{mk_{l}}}$. Finally,
the set of nonlinear differential equations is expressed in non-dimensional
form as
\begin{equation}
\ddot{u}_{n}+2\xi\dot{u}_{n}+u_{n}-\alpha(u_{n+1}+u_{n-1}-2u_{n})+u_{n}%
^{3}=g_{n}(\tau), \label{Eq:FinEqs}%
\end{equation}
where the dots represent the derivative with respect to the new time scale
$\tau=\omega_{0}t$, $\alpha=\frac{\omega_{c}^{2}}{\omega_{0}^{2}}$,
$A=\frac{\omega_{0}}{\sqrt{\gamma}}$, $g_{n}(\tau)=\frac{f_{n}(t)}{k_{l}A}$,
and $x_{n}(t)$=$Au(\tau)$. Moreover, one has the cyclic boundary condition
$u_{N_{s}+1}=u_{1}$ and $u_{0}=u_{n}.$

\section{Conservative regime: linear analysis}

The linear and conservative regime of Eq. \eqref{Eq:FinEqs}, for which the equation of
motion is given by:
\begin{equation}
\ddot{u}_{n}+u_{n}-\alpha(u_{n+1}+u_{n-1}-2u_{n})=0,\label{Eq:LinReg}%
\end{equation}
has been previously studied in the literature (see e.g. Refs.
\cite{Grolet2012}). The system has analytical {travelling wave} solutions such
that
\begin{equation}
u_{n}(\tau)=U_{n}^{k}\exp\{\mbox{i}\left[  k(n-1)a-\omega_{k}\tau\right]
\}+\mbox{c.c},\label{Eq:SolLin}%
\end{equation}
where $U^{k}$ is the complex wave amplitude, $a=2\pi/N_{s}$ is the sector
angle, $k$ is the wave number, $\omega_{k}$ is the natural angular frequency,
while c.c. states the complex-conjugate of the first expression. After
inserting Eq. \eqref{Eq:SolLin} into Eq. \eqref{Eq:LinReg}, it is possible to
demonstrate that the linear dispersion relation is given by
\begin{equation}
\omega_{k}^{2}=1+2\alpha\left[  1-\cos(ka)\right]  ,\label{Eq:DispRel}%
\end{equation}
which represents the natural angular frequencies of Eq. \eqref{Eq:LinReg}.
Moreover, the corresponding eigenvectors are obtained by substituting Eq.
\eqref{Eq:DispRel} into Eq. \eqref{Eq:SolLin}, and three different solutions
are possible: (1) if $k=0$ all masses move in phase such as $U_{n}^{k=0}=1$;
(2) if $N_{s}$ is even $k\in\lbrack1,\frac{N_{s}}{2}-1]$ (if $N_{s}$ is odd
$k\in\lbrack1,\frac{\left(  N_{s}-1\right)  }{2}-1]$) two sets of solutions
are possible such as $U_{n}^{k,cos}$=$\cos((n-1)ka)$ and $U_{n}^{k,sin}$%
=$\sin((n-1)ka)$; while (3) if $k=\frac{N_{s}}{2}$ the masses move out of
phase such as $U_{n}^{k=\frac{n}{2}}$=$(-1)^{n}$. In Fig. 2 the linear mode
shapes are shown assuming $N_{s}=6$. Notice that the mode shapes
(eigenvectors) do not depend on the coupling parameter $\alpha.$

\begin{center}
\includegraphics[
height=2.3949in,
width=5.0788in
]{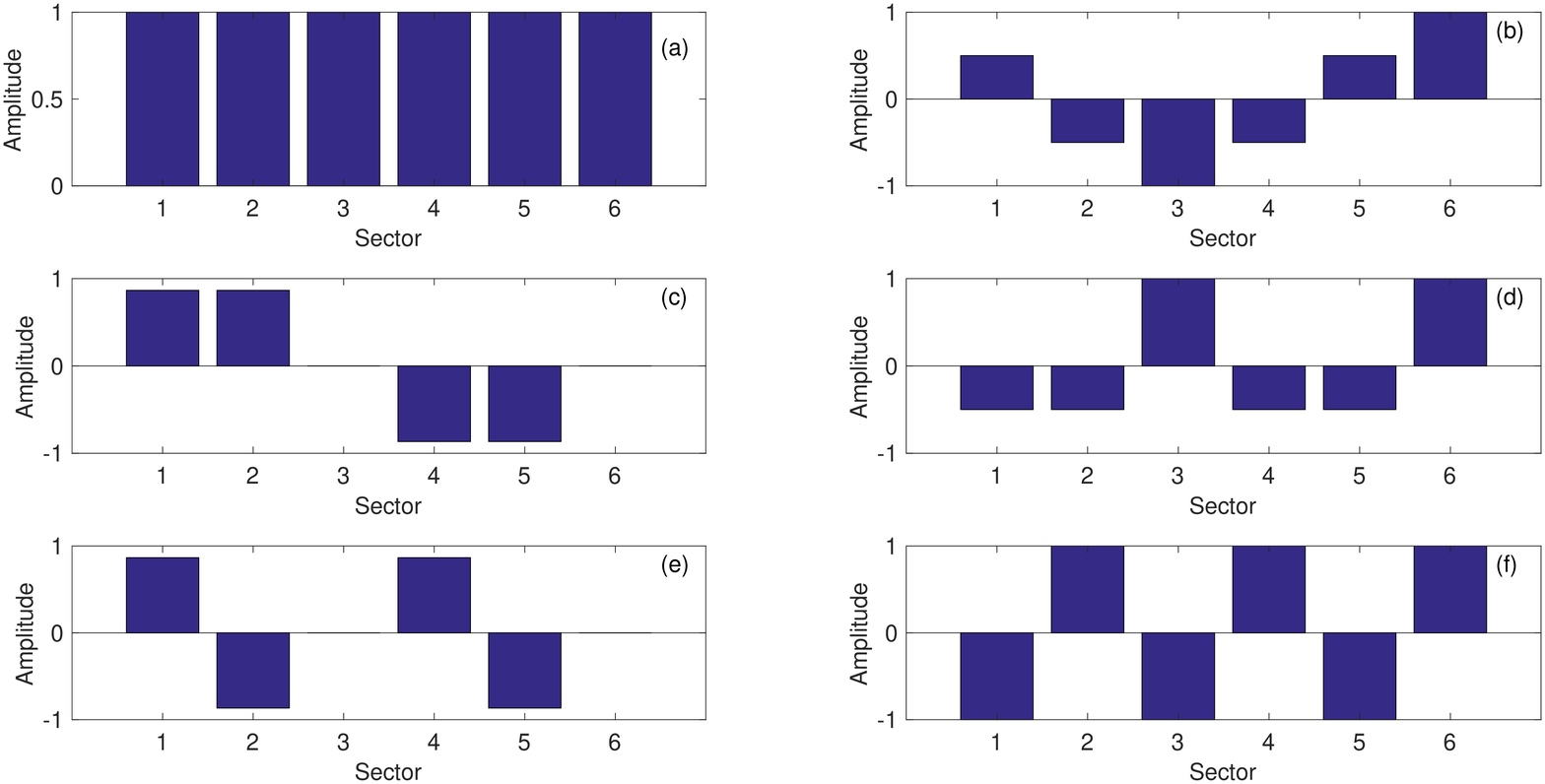}

Fig. 2 - Linear mode shapes of the system with $N_{s}=6$.
\end{center}

\section{Nonlinear normal modes}

\subsection{General analysis}

\label{Sub:NNM}

In the nonlinear and conservative regime our investigation is based on the
concept of nonlinear modes (\cite{Rosenberg},\cite{ShawPierre}). Therefore, periodic vibrations in the form of
\begin{equation}
\ddot{u}_{n}+u_{n}-\alpha(u_{n+1}+u_{n-1}-2u_{n})+u_{n}^{3}=0 \label{Eq:NlEqs}%
\end{equation}
are computed using only one harmonic such as
\begin{equation}
u_{n}=U_{n}\exp\{\omega\tau\}+c.c., \label{un}%
\end{equation}
where $U_{n}$ is the {complex-valued} amplitude and $\omega$ is the nonlinear
natural angular frequency. The previous expression (\ref{un}) is inserted into
Eq. \eqref{Eq:NlEqs} and the resulting equation {is} projected into the
Fourier basis which, neglecting higher order harmonics, leads to the following
set of algebraic equations
\begin{equation}
(1+2\alpha-\omega^{2})U_{n}+3U_{n}^{3}-\alpha(U_{n+1}+U_{n-1})=0.
\label{Eq:Alg_NNM}%
\end{equation}
The solutions of Eq. \eqref{Eq:Alg_NNM} define the nonlinear normal modes
(NNMs) of the physical system depicted in Fig. 1. Moreover, due to the
amplitude dependency of the natural angular frequencies, some of the nonlinear
mode shapes loose their homogeneity. Some modes can localize when the
amplitude increases, i.e. only a small subset of the oscillators are vibrating
with significant amplitudes.

The computation of the non-linear modes is carried out by solving
Eq.\eqref{Eq:Alg_NNM} using a continuation procedure. In this study we used
the numeric asymptotic method presented in (\cite{Cochelin1},\cite{Cochelin2}%
), and implemented in the MANLAB package. It is common to assume at low
amplitude the initial points of the continuation procedure are given by the
linear mode shapes, whereas other possible starting points at high amplitude
will be described later. The results of the continuation procedure provides
the evolution of the natural angular frequency with respect to the amplitude
of motion. One also obtains the evolution of mode shapes as the amplitude of
vibration increases (\cite{Rosenberg},\cite{ShawPierre}). From the results,
one can see that some mode shapes are similar (i.e. the shape does not depend
on the amplitude of motion), and some others are non-similar. Potential
bifurcation points of a given non-linear mode are computed searching for
points where the determinant of the Jacobian matrix vanishes. The Jacobian
matrix of Eq. \eqref{Eq:Alg_NNM} is given by
\begin{equation}
\textbf{J}=%
\begin{pmatrix}
\xi_{1} & -\alpha & 0 & \dots & 0 & -\alpha\\
-\alpha & \xi_{2} & -\alpha & 0 & \dots & 0\\
\dots &  &  &  &  & \\
0 & \dots & 0 & -\alpha & \xi_{N-1} & -\alpha\\
-\alpha & 0 & \dots & 0 & -\alpha & \xi_{N}\\
&  &  &  &  &
\end{pmatrix}
\text{ }\label{jacobian}%
\end{equation}
with $\xi_{n}=1+2\alpha-\omega^{2}+9U_{n}^{2}.$ At high amplitude the
nonlinear term will be dominant, hence the linear coupling between the
oscillators can be neglected, therefore the system is equivalent to a series
of \textit{uncoupled} Duffing oscillators (for $n\in\lbrack1,N]$):
\begin{equation}
\ddot{u}_{n}+u_{n}+u_{n}^{3}=0
\end{equation}

The eigen-shapes of such a system are given by the vectors of the canonical
basis of $\mathbb{R}^{N}$, and the evolution of the angular frequency (for
oscillator $n$) using a single harmonic HBM is given by $\omega_{n}%
^{2}=1+3U_{n}^{2}$.

In the case of a non-linear modes motion, all (uncoupled) oscillators must
have the same frequency, this is possible only if the amplitude are the same
for all oscillators ($U_{n}=U_{0}$ or $U_{n}=-U_{0}$) or if some amplitudes
are zero. Therefore, at high amplitude, one expects that the mode shape will
have the following general form:
\begin{equation}
\phi_{n}=u_{n},\text{ with }u_{n}\in\{-1,\ 0,\ 1\}
\end{equation}
In other words, there are three choices for each component of the mode shape.
This leads to $3^{N_{s}}$ different mode shapes at high amplitude. Those mode
shapes can be used as starting points for relatively high amplitude of motion
to initiate the continuation procedure, which is carried out backward, i.e. in
the direction of decreasing frequency. However, due to the symmetry of the
system, several of those shapes belong to the same "family", i.e. they differ
only by a rotation or by a change of sign (see Appendix - A). Without loss of
generality, in the following we will choose only one representative for each
\textquotedblright family\textquotedblright\ to drastically decrease the
number of possible initial points.

\subsection{Nonlinear normal modes analysis for the present system}

We start our investigation focusing on periodic solutions from the undamped
and unforced nonlinear system (\ref{Eq:Alg_NNM}) assuming $N_{s}=6$ and
$\alpha=0.01$. Therefore, the strategy proposed in Sub. \ref{Sub:NNM} is
employed to compute the NNMs of the underling conservative regime. The details
of the derivation are not reported, while the main results are highlighted here:

\begin{itemize}
\item the first NNM {$U_{n}^{k=0}$}, when all masses vibrate in phase, is
similar (its shape does not change with amplitude) and does not bifurcate;

\item the NNM $U_{n}^{k=1,cos}$ is non-similar and bifurcate twice;

\item the NNM starting from $U_{n}^{k=1,sin}$ is similar and does not bifurcate;

\item the NNM starting from $U_{n}^{k=2,cos}$ is similar and does not bifurcate;

\item the NNM starting from $U_{n}^{k=2,sin}$ is non-similar and bifurcates twice;

\item the last NNM {$U_{n}^{k=3}$}, when all masses vibrate out of phase, is
similar and bifurcates eleven times.
\end{itemize}

Starting from the linear mode, and following the bifurcations, $22$ classes of
solutions have been computed (thin solid lines in Fig. 3). From symmetry
consideration the number of possible solutions can be reduced from $3^{N_{s}%
}=729$ to $52$ (including the trivial solution, see Appendix - A) $22$ of
which have been computed continuing the linear modes. In order to compute the
remaining $30$ modes, we use the corresponding shapes as a starting point at
high amplitude for the continuation algorithm, and we start the continuation
"backward" (i.e. in the direction of decreasing frequency). This allows to
compute new branches of solution in the Frequency Energy Plot (FEP) diagram (Fig. 3), where the
$L^{2}$ norm is defined as%

\begin{equation}
L^{2}=\sqrt{\sum_{i=1}^{N_{s}}U_{n}^{2}}, \label{Eq:L2norm}%
\end{equation}

The non-linear modes corresponding to those new branches are not connected to
the other modes through bifurcation and appear as isolated branches on the FEP
diagram (dashed thick red lines in Fig. 3). Those mode{s} do no exist at low
amplitude, but appear only after the energy is greater than a given threshold
(depending on the mode). At high energy the nonlinear term is dominant and the
oscillators behave as isolated. In fact the corresponding NNMs cluster into 6
groups, which energy depends on the number of oscillators vibrating with no
negligible amplitude.

Fig. 4 reports some NNMs shapes (we have selected 15 out of the 52 solutions). Notice
that possible solutions involve high localized ones with very few oscillators
vibrating with appreciable amplitude.

\begin{center}
\includegraphics[
height=3.0877in,
width=5.0788in
]{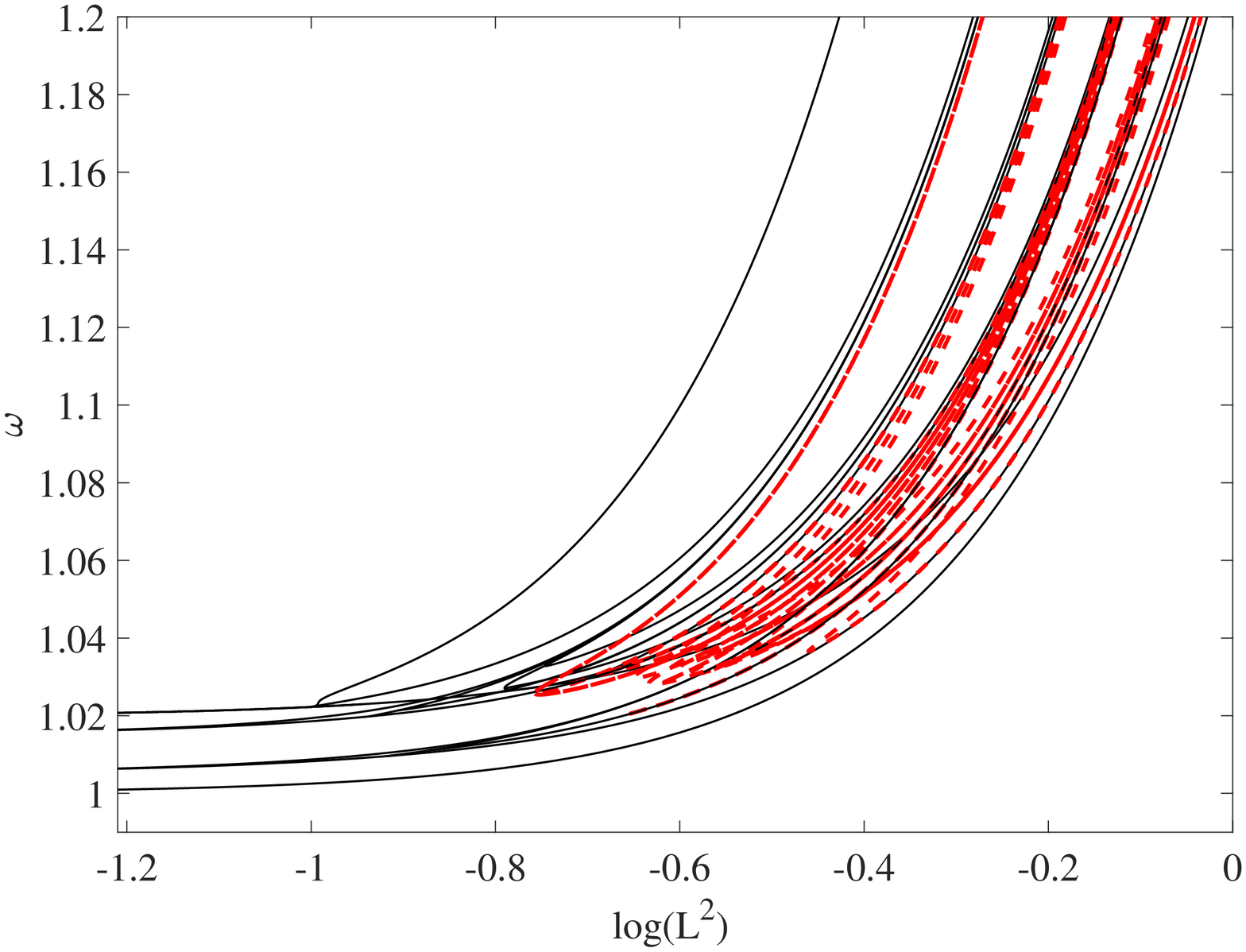}

Fig. 3 - FEP diagram of the nonlinear modes with their bifurcation. Thin solid black
lines represent the solution of \eqref{Eq:Alg_NNM} starting, at low energy,
from the linear modes. Thick dashed red lines represent the solution of
\eqref{Eq:Alg_NNM} starting at high energy and continuing them towards lower frequency (see Appendix).

\includegraphics[
height=4.6141in,
width=5.0788in
]{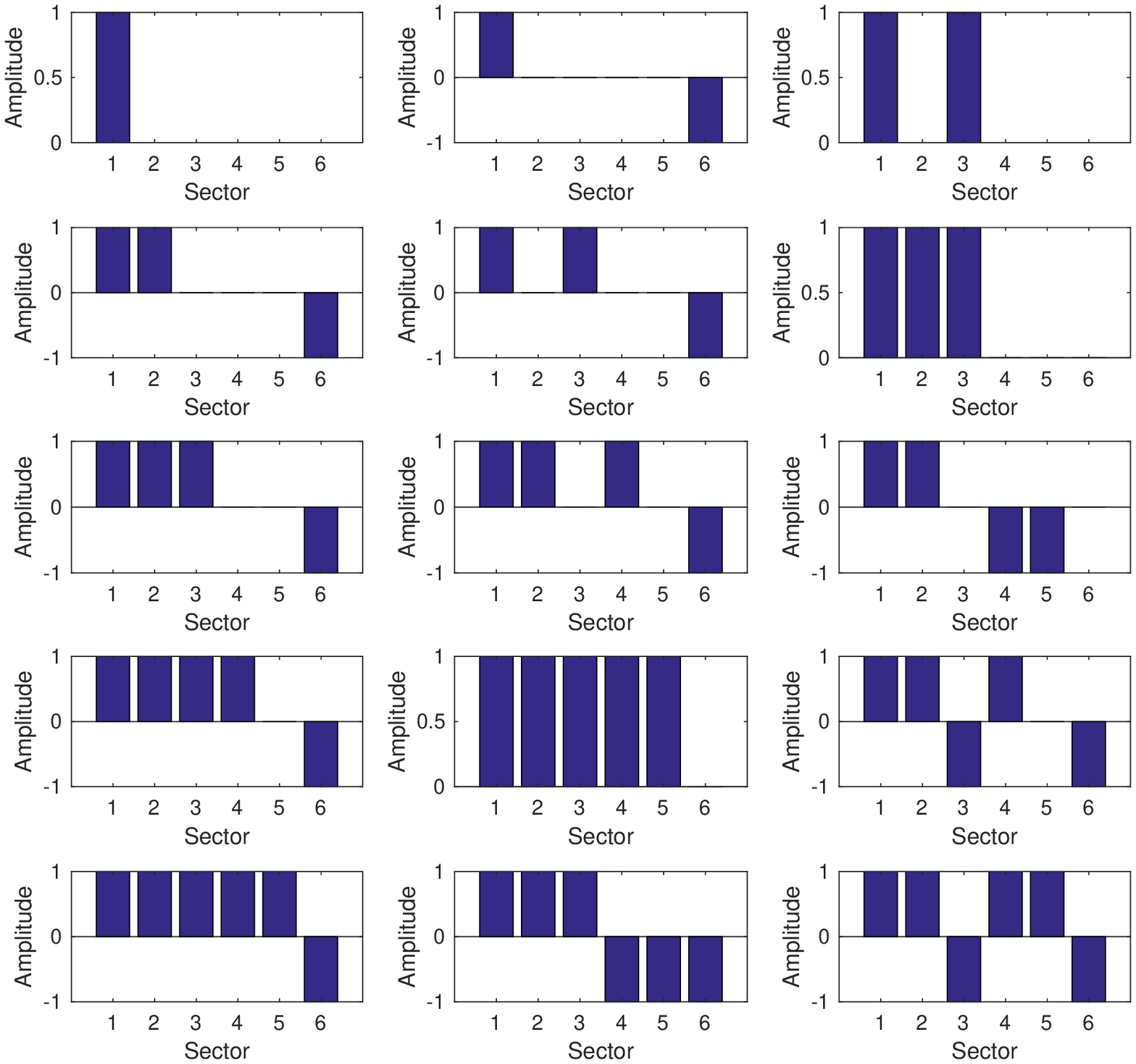}

Fig. 4 - A selection of 15 (out of 52) nonlinear normal mode shapes with
different degree of localization
\end{center}

\section{Nonlinear forced and damped regime}

\subsection{Solution approach}

In the present analysis we investigate the system described by Eq.
\eqref{Eq:FinEqs} subjected to an external force%
\begin{equation}
g_{n}(\tau)=\tilde{G}_{n}\exp\{\mbox{i}\omega_{F}\tau\}+\mbox{c.c.},
\label{Eq:Force}%
\end{equation}
where $\omega_{F}$ is the external force angular frequency and $\tilde{G}_{n}$
is the force amplitude. To find the steady state periodic solutions of the
system we adopt the harmonic balance method. Assume that, in the weakly
nonlinear regime, the response $u_{n}(\tau)$ is well approximated by a single
harmonic such as
\begin{equation}
u_{n}(\tau)=\tilde{U}_{n}\exp\{\mbox{i}\omega_{F}\tau\}+\mbox{c.c.},
\label{Eq:Resp}%
\end{equation}
where $\tilde{U}_{n}$ is a constant amplitude. After inserting Eqs.
\eqref{Eq:Force} and \eqref{Eq:Resp} into Eq. \eqref{Eq:FinEqs} and projecting
onto the truncated Fourier basis, the following set of nonlinear algebraic
equations is obtained
\begin{equation}
(-\omega_{F}^{2}+2\mbox{i}\xi\omega_{F}+1+2\alpha)\tilde{U}_{n}-\alpha
(\tilde{U}_{n+1}+\tilde{U}_{n-1})+3|\tilde{U}_{n}|^{2}\tilde{U_{n}}=G_{n}
\label{Eq:Forced}%
\end{equation}
is obtained. Therefore, the amplitudes and phases of $u_{n}(\tau)$ are defined
by the solutions of the algebraic system in Eq. \eqref{Eq:Forced}.

\subsection{Nonlinear forced responses}

In this subsection we simulate the system in Fig. 1, as described by the
non-dimensional form in Eq. \eqref{Eq:FinEqs}, assuming $N_{s}=6$ masses,
$\xi=0.005$ and weak coupling condition $\alpha=0.01$. A harmonic force is
applied to each oscillator assuming a constant amplitude and phase value, such
as
\begin{equation}
g(\tau)=G_{0}\exp\{\mbox{i}\omega_{F}\tau\}+\mbox{c.c.},
\end{equation}
where $G_{0}$ is the force amplitude. One should note that the excitation is
perfectly orthogonal to all linear normal modes, except the first one which
represents vibrations with all masses moving in phase. However, due to the
presence of damping, the responses may vibrate with different phase values.
For this purpose we used the continuation package AUTO, using as initial conditions highly loacalized guess for the vibration shape (e.g. [0,1,0,0,-1,0])

The system frequency response function ($L^{2}$-norm vs angular frequency
$\omega_{F}$) is depicted in Fig. 5.

\begin{center}
\includegraphics[
height=3.0877in,
width=5.0788in
]{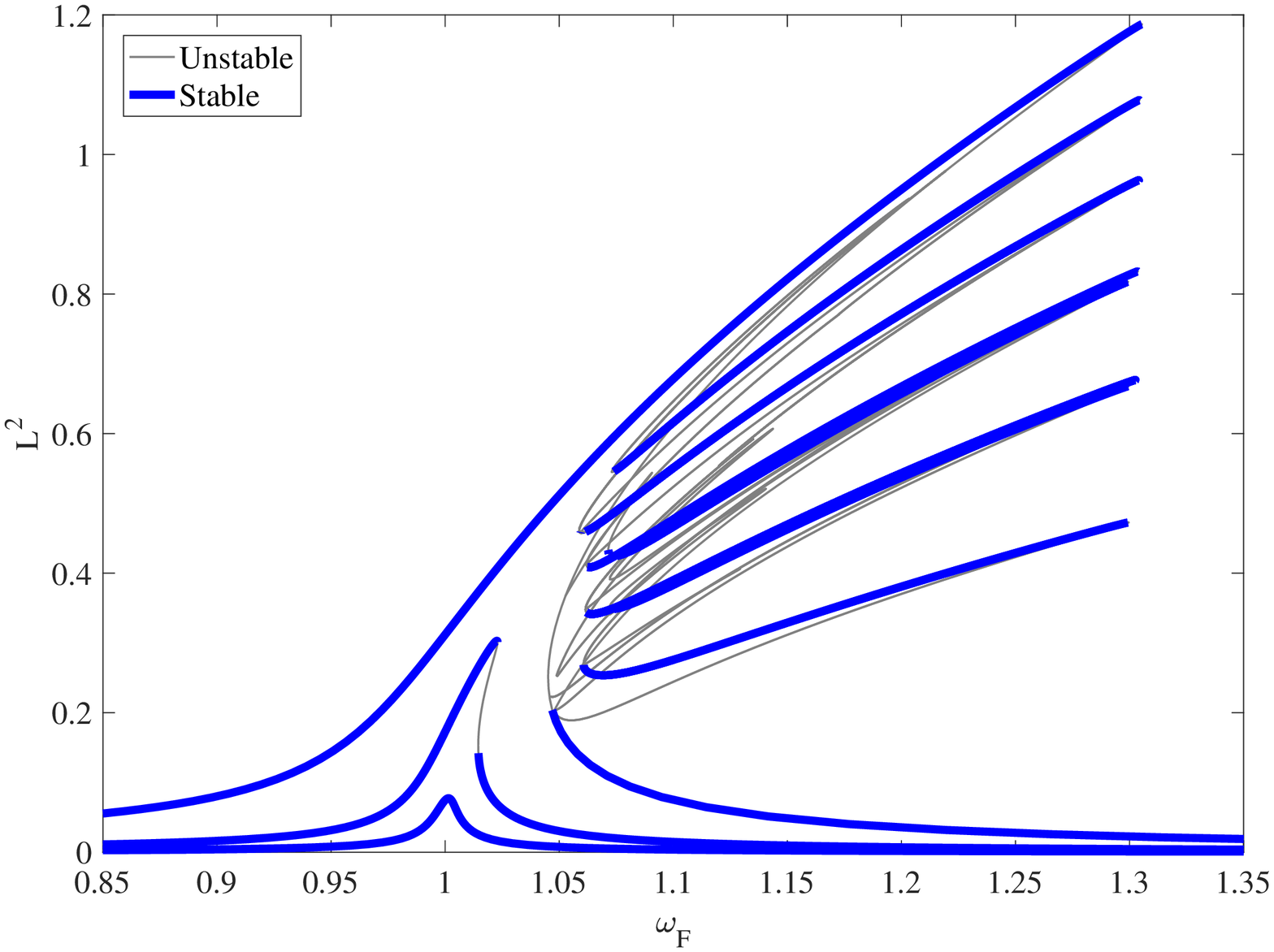}

Fig. 5 - Frequency response function obtained for the chain of weakly coupled
Duffing oscillators computed for three different force levels $G_{0}=[0.32,
1.3, 6.3]\times10^{-3}$
\end{center}

The results were computed using three different force levels, while blue thick
lines indicate linearly stable solutions. The first response, computed for
$G_{0}=3.2\times10^{-4}$, illustrates the response when the nonlinear force is
negligible. The response is linear and single-valued. The second curve,
computed for $G_{0}=1.3\times10^{-3}$, shows already the stiffening effect
induced by the cubic stiffness. However, the responses are still homogeneous
and no localization is observed. In the last case, when $G_{0}=6.3\times
10^{-3}$, the response shows several bifurcated branches. The forced response shows multiple solutions within the frequency interval where the forced single oscillators exhibit two stable solutions for a given exciting frequency.
Hence the forced response of the whole system is multivalued with many intertwining
branches. The new branches seem to organize themselves similarly to snaking
bifurcations (\cite{Hunt,Hunt2000,Burke},\cite{Champneys1998}). In fact they
fold alternating each individual oscillator from an upper branch to a lower
one. Therefore, it is possible to observe solutions varying from states where
energy is localized in just one oscillator, to solutions when energy is
equally spread along the structure. This pattern is not much affected by the particular choice of cyclic (which we are using here) or free-free boundary conditions. Notice that, although the branches seem to
be all connected, in fact they aren't, but appear as isolas of solutions,
which get superposed in a two dimensional representation. 

In order to provide a closer look at the transition from homogeneous states to
localized ones, eq. \eqref{Eq:FinEqs} is solved for four different excitation
levels. The results are depicted in Fig. 6 increasing the exciting force from
panel (a) to panel (d) respectively. On the left column the frequency response
function is reported, while, on the right column, the modulus of the vibration
amplitude is shown for the labeled solutions. Linearly stable solutions are
drawn with thick blue lines, unstable solutions with thin gray lines. Results
in panel (a) have been computed for $G_{0}=1.6\times10^{-3}$. It is shown that
for low excitation the response is quite homogeneous (Fig. 6, panel (a), right
column) and only the unstable solution labeled with \textquotedblright
P4\textquotedblright\ shows some degree of localization. When the force is
slightly increased to $G_{0}=2.4\times10^{-3}$ (Panel (b)) other branches of
solutions are identified. Moreover, spatially localized linearly stable states
appear. The bifurcation picture is mainly composed of two localized states
detaching from the homogeneous solutions (P1 and P2), and two isolas that do
not bifurcate from the main solution (solutions P3 and P4). Panel (c) and (d)
are computed for $G_{0}=3.2\times10^{-3}$ and $G_{0}=6.4\times10^{-3}$
respectively. It is shown that increasing the external excitation more
branches are identified, and the isolas seem to evolve towards a final states
where they merge with the upper branch of localized solution. Finally the
final \textquotedblright snaking bifurcation pattern\textquotedblright%
\ appears where the vibration gets localized from one single oscillator to the
final homogeneous configuration either in the upper or in the lower branches. Comparing Fig. 4 and Fig. 6 one also notice the similarities between the NNM and forced responses shapes. Indeed, this behavior should be expected, as it is well known that the NNMs are the "backbone curves" of the forced response and, at the resonance point, the forced system behaves alike the underlying NNM.

\begin{center}
\includegraphics[
width=6.5in]{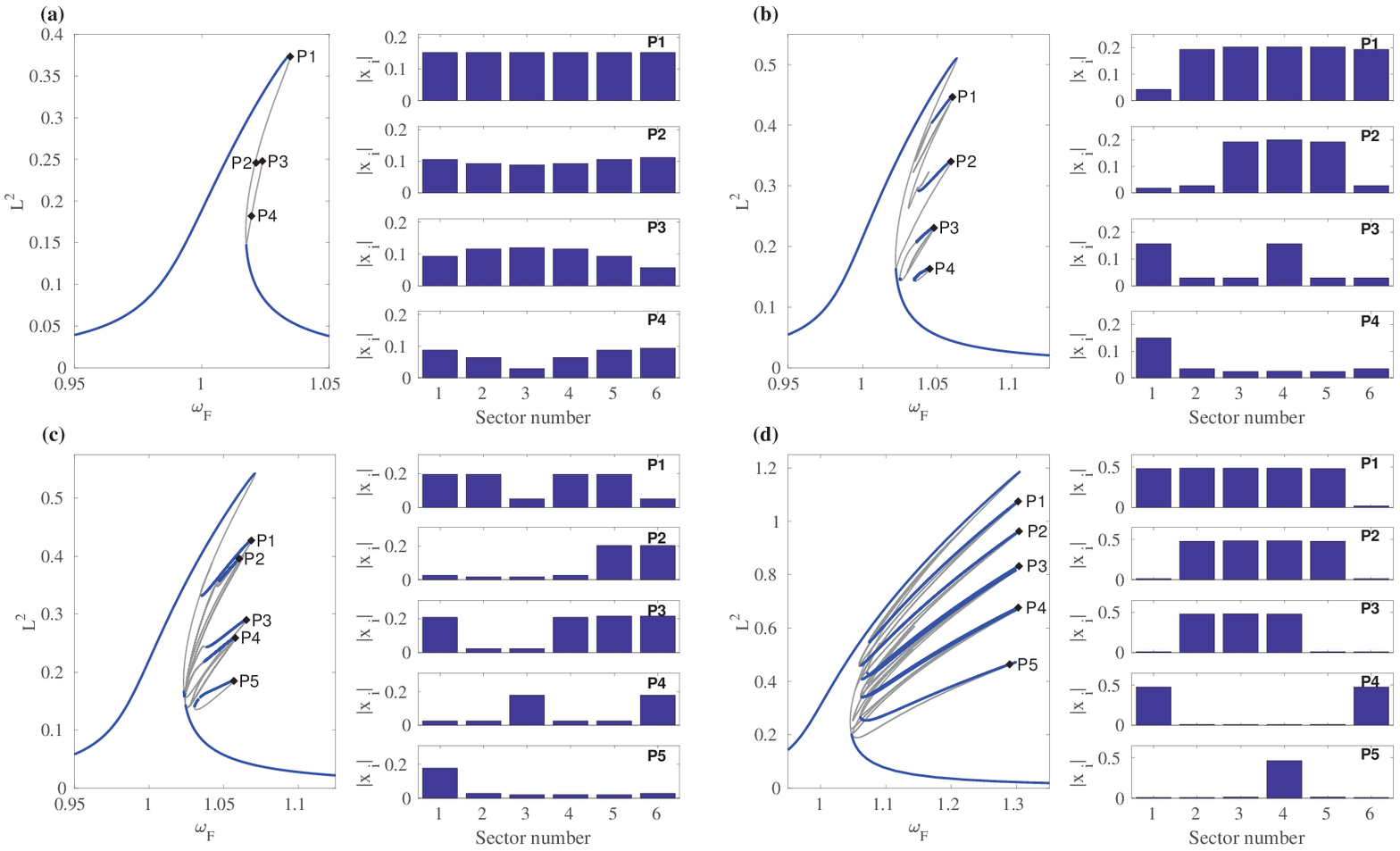}
\end{center}

\begin{center}
Fig. 6 - Form panel (a) to panel (d) the excitation force is $G_{0}=[1.6, 2.4,
3.2, 6.4]\times10^{-}3$ respectively. For each panel on the left column: frequency response function
(blue thick lines for linearly stable solutions, thin gray lines for unstable
solutions). For each panel on the right column: vibration amplitude (in modulus) for each sector
\end{center}

In Fig. 7 the nonlinear normal modes of the dynamical system (thin dashed
lines) are superposed on the forced response (only stable solutions are
reported with thick solid lines) for $G_{0}=6.3\times10^{-3}$. It is clearly
shown that the local resonances of the mechanical systems are associated to a
particular backbone curve. At high energy the response clearly clusters into 6
groups which correspond to the number of oscillators involved in the vibration.

\begin{center}
\includegraphics[
height=3.0877in,
width=5.0788in
]{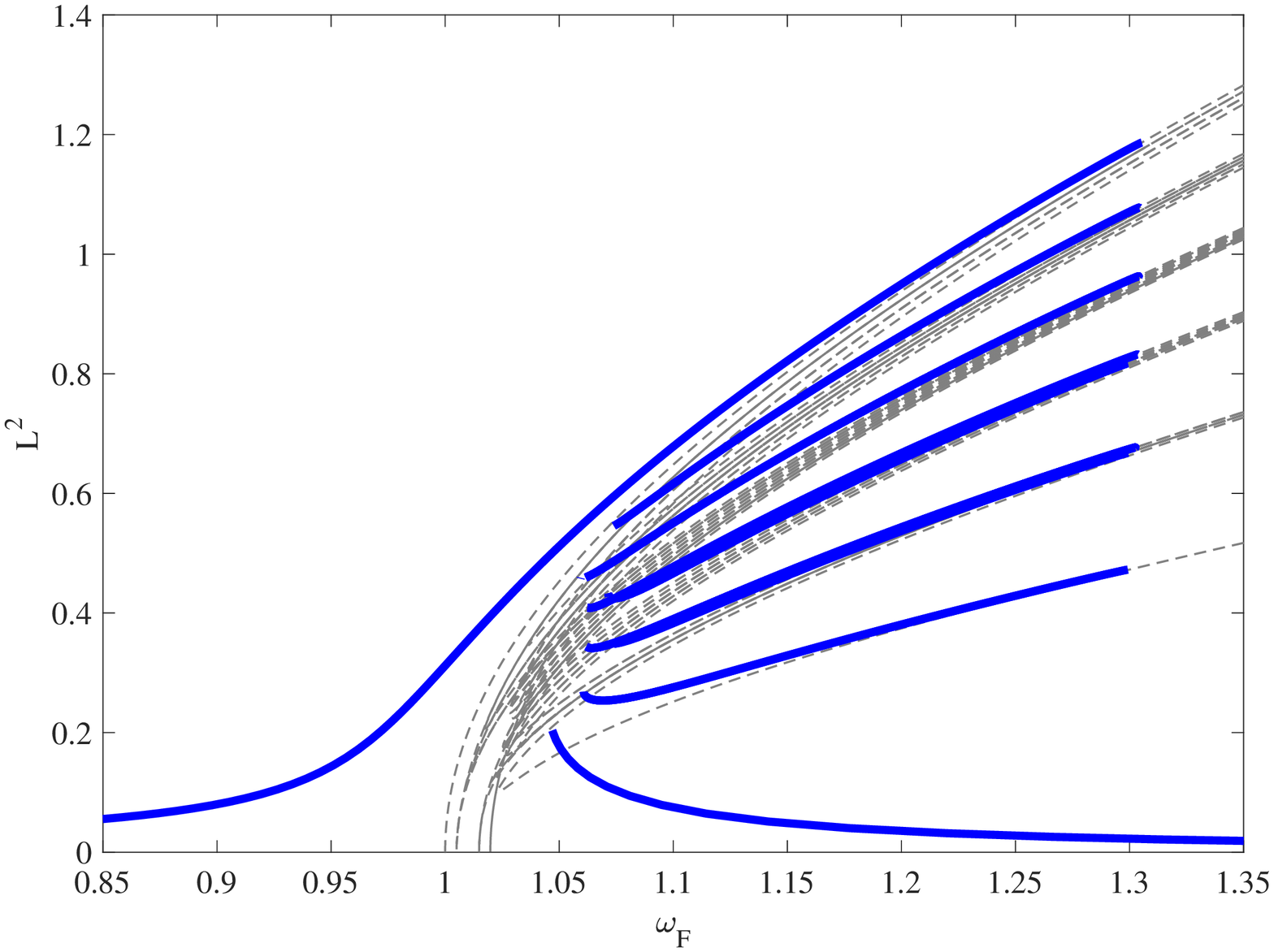}

Fig. 7 - The frequency response function of the mechanical system for
$G_{0}=6.3\times10^{-3}$ is superposed to the NNMs. Only the stable solutions
of the forced response are reported, while dashed lines represent the NNMs.
\end{center}

\section{Conclusions}

In this paper, the problem of vibration localization in externally excited weakly-coupled cyclic symmetric structures affected by geometrical nonlinearities has been considered. The model we have set-up is
constituted by a cyclic symmetric chain of six Duffing oscillators, where
stiffening effects have been considered via a cubic stiffness term. The
nonlinear normal modes have been computed. Due to symmetry
considerations, the system periodic solutions can be classified into $52$
different shapes (including the trivial one). It has been shown that $22$ of
them are obtained starting from the linear modes and following their
bifurcation from low to high amplitude. The remaining $30$ solutions, instead,
appear to be isolated branches, detached from the other modes. The isolated
branches exist only after a certain energy threshold is reached and disappear
at low energy.

When the cyclic symmetric chain is forced with the first linear mode shape
(all the oscillators are excited with the same force) spatial localized
vibrating states appear. The degree of localization is strongly dependent on
the excitation strength. For strong excitation more localized branches appear,
which, in the bifurcation diagram arrange similarly to snaking bifurcations.
When the backbone curves of the nonlinear normal modes are superposed with the
forced response they appear to match almost exactly, showing that the
intermediate states between the low and high energy states correspond to local
resonances of the system. We have shown that this behaviour is strongly
related to the bistable behaviour of the single nonlinear oscillators from which the chain is formed: if the excitation level is low, the bistable range
disappears, the system behaves nearly linearly and there is no localization. In the latter case, there is no localization.

The results of the present study may be of interest for aerospace/aeronautical
applications, where the quest of high performance is leading to lightweight and
slender structures, often characterized by weak nearest
neighbor coupling (e.g. fan blades or antennas). Due to high flexibility, the
latter are prone to large oscillations thus to nonlinear stiffening effects.
This work shows that vibration localization in space can arise due to nonlinear phenomena, which represents a different mechanism with respect to mistuning. Further studies are needed to
better understand how the presently demonstrated nonlinearly triggered localization effect interacts with linear localization phenomena such us mistuning.

\setcounter{secnumdepth}{0}

\section{Author Contribution Statement}

AP and FF conceived the work. FF computed the numerical results. AG did the
NNM analysis. AP wrote the paper and coordinated all the work. MC and NH supervised the work.

\setcounter{secnumdepth}{0}

\section{Acknowledgements}

AP is thankful to the DFG (German Research Foundation) for funding the
projects PA 3303/1-1 and HO 3852/11-1.

\setcounter{secnumdepth}{0}

\section{Appendix A: sorting initial points using symmetry}

Denote $E$ the set of all possible initial points at high amplitude. There are
$|E|=$card$\left(  E\right)  =3^{N_{s}}$ possible choices, i.e. 3 choices
$\left[  1,0,-1\right]  $ for the $N_{s}$ components.

In order to reduce the number of initial points, one can use the fact that the
algebraic system defining the non-linear modes in Eq. (\ref{Eq:Alg_NNM}) is
equivariant under several groups:

\begin{itemize}
\item The cyclic group $\mathcal{C}_{N}$, which representation in a $N$
dimensional space is generated by the $N\times N$ matrix $\mathbf{R}$ given by
the following:
\begin{equation}
\mathbf{R}=%
\begin{pmatrix}
0 & 1 & 0 & 0 & 0 & 0\\
0 & 0 & 1 & 0 & 0 & 0\\
\vdots &  &  & \ddots &  & \vdots\\
0 & 0 & 0 & 0 & 1 & 0\\
0 & 0 & 0 & 0 & 0 & 1\\
1 & 0 & 0 & 0 & 0 & 0\\
&  &  &  &  &
\end{pmatrix}
\end{equation}

\item The axial symmetry, generated by the $N \times N$ matrix $\mathbf{S}$
given as:
\begin{equation}
\mathbf{S} =
\begin{pmatrix}
1 & 0 & 0 & 0 & 0 & 0\\
0 & 0 & 0 & 0 & 0 & 1\\
0 & 0 & 0 & 0 & 1 & 0\\
\vdots &  &  & {\mathstrut^{.^{.^{.}}}} &  & \vdots\\
0 & 0 & 1 & 0 & 0 & 0\\
0 & 1 & 0 & 0 & 0 & 0\\
&  &  &  &  &
\end{pmatrix}
\end{equation}

\item The "sign-change" symmetry (related to the fact that we are looking for
"symmetric" periodic solution such that $u(t+\frac{T}{2})=-u(t)$), generated
by the $N\times N$ matrix $\mathbf{C}$ given as:
\begin{equation}
\mathbf{C}=%
\begin{pmatrix}
-1 & 0 & 0 & 0 & 0\\
0 & -1 & 0 & 0 & 0\\
\vdots &  & \ddots &  & \vdots\\
0 & 0 & 0 & -1 & 0\\
0 & 0 & 0 & 0 & -1\\
&  &  &  &
\end{pmatrix}
\end{equation}

\end{itemize}

To sum up the algebraic system is equivariant under a group $G$, having
$|G|=4N$ elements, which can be represented as:
\begin{equation}
G=\{g_{ijk}=\mathbf{S}^{i}\mathbf{C}^{j}\mathbf{R}^{k},\text{ with
}(i,\ j,\ k)\in\lbrack0,\ 1]\times\lbrack0,\ 1]\times\lbrack0,\ N-1]\}
\end{equation}
For the case $N_{s}=6$ taking into account the symmetry conditions, the number
of initial points reduces from $|E|=3^{N_{s}}=729$ to only $52$, which
includes the trivial one with all $0$.

\end{document}